\documentclass[epj]{webofc}
\pdfoutput=1
\usepackage[utf8]{inputenc}
\usepackage[varg]{txfonts}   
\usepackage{booktabs}
\usepackage{xcolor}
\definecolor{darkred}{rgb}{0.4,0.0,0.0}
\definecolor{darkgreen}{rgb}{0.0,0.4,0.0}
\definecolor{darkblue}{rgb}{0.0,0.0,0.4}
\usepackage{subfigure}
\wocname{EPJ Web of Conferences}
\woctitle{Lattice2017}
\newcommand{\bea}{\begin{eqnarray}}
\newcommand{\ena}{\end{eqnarray}}

\newcommand{\dlangle}{\left\langle \kern-.17em \left\langle}
\newcommand{\drangle}{\right\rangle \kern-.17em \right\rangle}

\begin{document}
%
\selectlanguage{english}
\rightline{CERN-TH-2017-231}
\title{%
Lattice QCD on new chips: a community summary
}
\author{%
\firstname{Antonio} \lastname{Rago}\inst{1,2}\fnsep\thanks{Speaker,
  \email{antonio.rago@plymouth.ac.uk}}
}
\institute{%
Centre for Mathematical Sciences, Plymouth University, 
Plymouth, PL4 8AA, United Kingdom
\and
CERN, Theoretical Physics Department,
Geneva, Switzerland
}
\abstract{%
I review the most recent evolutions of the QCD codes on new
architectures, with a focus on the performances obtained by the different
coding strategies as presented during the Lattice2017 conference.
}
\maketitle
\section{Introduction}\label{intro}
\begin{figure}[thb] 
  \centering
\includegraphics[width=.5\textwidth]{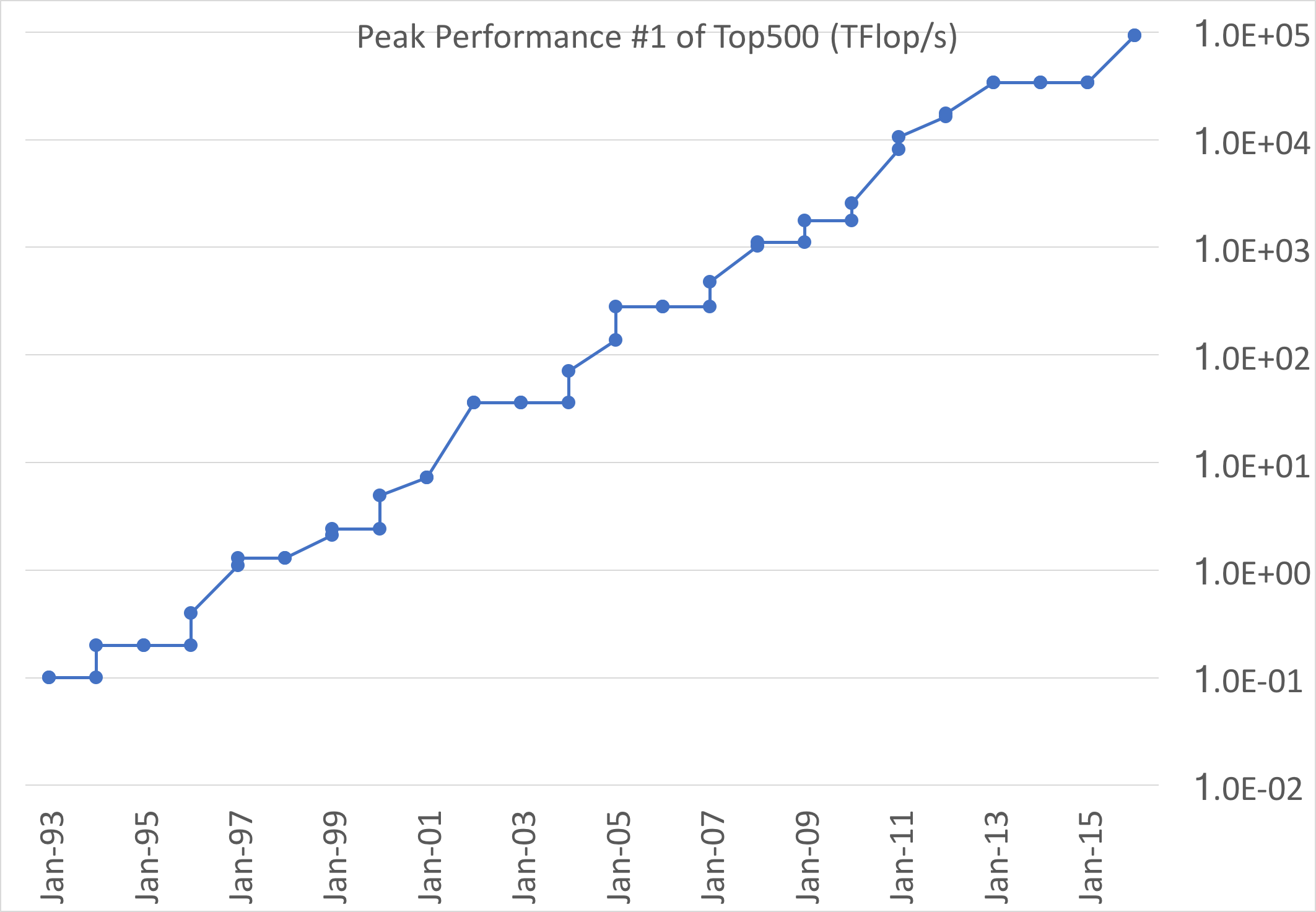}
\caption{Peak performance of the top supercomputer of the world over
  the years (data from TOP500 \cite{top500}).}
  \label{moore}
\end{figure}

In 1965 Gordon Moore formulated his famed law: companies should be
capable of doubling the
number of transistors on a given area of chip every two years \cite{moore:1965}. The law
has roughly held true up to now, with the caveat that the original
formulation has been enlarged to include other aspects than raw
single-CPU frequency scaling (see figure \ref{moore}). Nowadays
Moore's law is attained by the increase in the
number of cores, larger vectorization units, and more performing memory
structure and hierarchy. 
Unfortunately the interconnection speed has
not kept up with the pace of progress of the other components.
You may also find various claims that we are getting close to the end of
the silicon era, with evolutions such as tunneling transistors and
spintronic. The impending doom has been announced at least ten years
ago and Moore's law is still holding, silicon-based. 

This hardware complexity and variety reflects into challenges for the
programmer who intends to write optimised code. While in the past only
scalar tuning needed to be considered to reach optimal performance,
nowadays the roadmap for the HPC programmer is definitely more
complex: to fully exploit the computational power of modern hardware, the programmer must think of
coordination and communication, threading (MIMD) and vectorisation
(SIMD). And on top of that careful data motion and memory and caching
strategies need to be orchestrated too. Each of these aspects will
be associated to hardware bounds (cache, memory, network \dots) and
the challenge programmers must face, to exploit full hardware
capabilities, is to saturate all these bounds together \cite{Boyle:2017vhi}. 

\section{Hardware and QCD applications}\label{sec-1}
In the following I will present the efforts of the lattice community
in developing code for the newest available hardware, as presented
during the Lattice 2017 conference.  

\subsection{Intel Xeon Phi}\label{knl}
The Xeon Phi name identifies a set of many core processors (MIC-many
integrated core) developed
by Intel and designed specifically for HPC purposes.
One of the principal features of this chipset is the full x86
compatibility, i.e. the capability of running software targeted at
standard x86 CPU. Another important feature is the compatibility with standard programming API such as OpenMP.
These features should facilitate code portability from other similar
architectures.

The Xeon Phi was initially developed as a PCIe coprocessors, with the
codename Knight Corner (KNC), while for the second generation it was
developed also as standalone CPU structure with codename Knights
 Landing (KNL).

The specific structure of the KNL  has been presented and discussed in many
publications \cite{jeffers2016intel,colfax,Boyle:2017vhi}, in this work we only highlight the
most relevant features for lattice simulations.

\subsubsection{KNL organisation}
\begin{figure}[thb] 
\hfill
\subfigure[Schematic diagram of the on-chip tiles arrangement on a processor.]{\includegraphics[width=.5\textwidth]{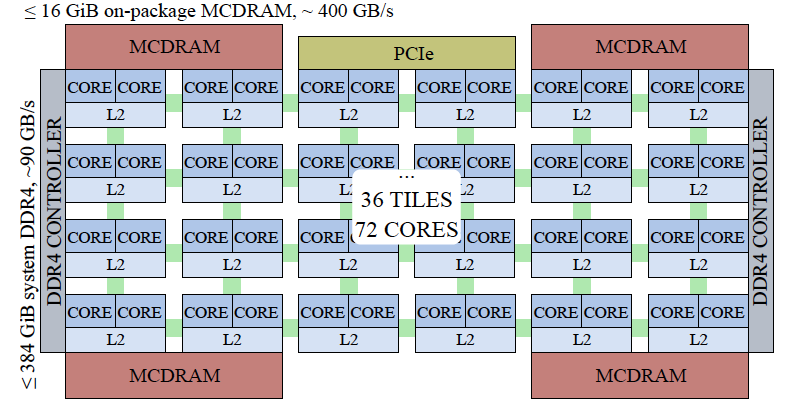}}
\hfill
\subfigure[Single tile structure.]{\includegraphics[width=.4\textwidth]{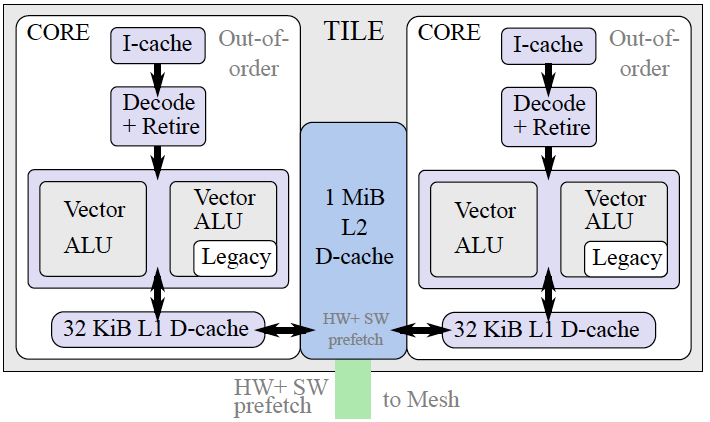}}
\hfill
\caption{KNL die organisation.}
  \label{KNL1}
\end{figure}

The structure of the KNL is reported in figure \ref{KNL1}.

It consists of a set of tiles with up to 36 tiles for KNL.
Each tile is equipped with 2 physical cores, for a maximum of 72
cores.
Each core shares a L2 cache on the tile that is
then  distributed with interconnection mesh as per figure \ref{KNL1}.a.
Each core is optimised for 4-way hyper-threading, hence up to
$4\times72=288$ logical processes per KNL.
The KNL is equipped with an on-package high-bandwidth memory MCDRAM
(16GB at$\sim475$GB/s \cite{stream}) connected by four controllers to the
processing cores.

\subsubsection{KNL CPU \& VPU}
The KNL tile is organised as per figure \ref{KNL1}.b.
The structure of KNL tile is designed for an hierarchical coding
paradigm with each core running multiple threads/processes (MIMD)
 and each thread (process) issuing vector instructions (SIMD). 
Each core has two Vector Processing Units (VPUs) and the out-of-order
pipeline is capable of two instructions per cycle. 
The optimal achievable benefit due to the vectorisation is:
\begin{itemize}
\item Single Precision $\to$  512 bit registers /32 bits $\times$ 2 VPUs = 32 SIMD lanes. 
\item Double precision $\to$ 512 bit registers /64 bits $\times$ 2 VPUs = 16 SIMD lanes. 
\end{itemize}
\begin{figure}[thb] 
\centerline{\includegraphics[width=.8\textwidth]{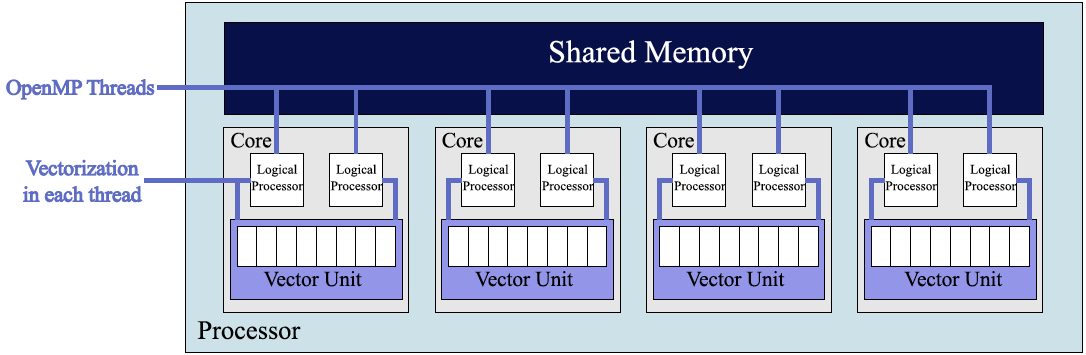}}
\caption{KNL coding hierarchy: co-existance with vectors.}
  \label{KNL2}
\end{figure}
\subsubsection{Key points in KNL optimisation}
\begin{figure}[thb] 
  \centering
 \hfill
   \subfigure[Flat mode.]{\includegraphics[height=.3\textwidth]{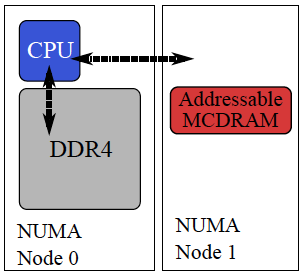}}
  \hfill
  \subfigure[Cache mode.]{\includegraphics[height=.3\textwidth]{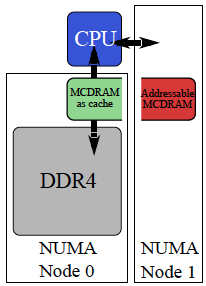}}
  \hfill
  \subfigure[Hybrid mode.]{\includegraphics[height=.3\textwidth]{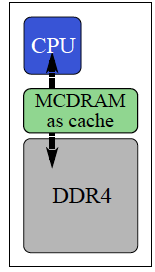}}
  \hfill
  \caption{KNL MCDRAM memory modes.}
  \label{KNL4}
\end{figure}

While designing a code for the KNL architecture, one has
to take into account the following key aspects:
\begin{itemize}
\item Vectorisation: Take full advantage of the 512 bit registers of
  the 2 VPU. Specifically noting that the legacy instructions (256
  bits or less) are
  available only on one of the VPU, it could even be beneficial to take
  advantage of the compiler automatic
  vectorisation.
\item Distributed L2 Cache with a mesh interconnect: data locality is
  of paramount importance. The locality can be exploited through three different
clustering modes characterised by a different degree of affinity
between distributed Tag Directory (TD) and cache. 

The three modes are:
\begin{itemize}
\item All-to-All: In this mode there is no affinity
between the TD and the memory (only for debug purpose).
\item Quadrant/Hemisphere: In this mode the TD and the memory reside
  in the same quadrant/hemisphere, see figure \ref{KNL3}.a.
The partitioning of the chip into quadrants is not visible
to the operating system, hence it is sufficient to maintain
data access locality to take full advantage of this mode.
\item SNC-4/SNC-2: In this mode the cores appear as 4 (or 2) NUMA
  nodes exposed to the operating system. In this case it
  is recommended to use a code that is NUMA aware, ensuring that each
  tile works mostly with its local memory, see figure \ref{KNL3}.b.
\end{itemize}
\begin{figure}[thb] 
  \centering
 \hfill
   \subfigure[Quadrant mode: Tag Directory and memory reside in the same quadrant.]{\includegraphics[width=.4\textwidth]{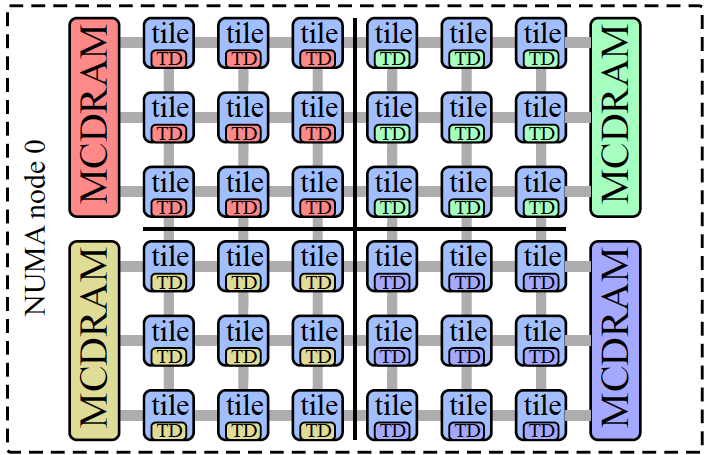}}
  \hfill
  \subfigure[SNC-4 mode: Cores appear as 4 NUMA nodes.]{\includegraphics[width=.4\textwidth]{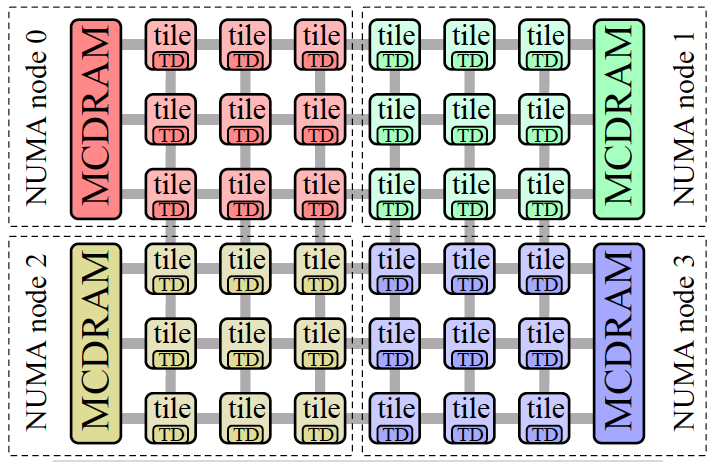}}
  \hfill
  \caption{KNL clustering modes.}
  \label{KNL3}
\end{figure}

\item
The MCDRAM has a bandwidth roughly 4 times bigger than the DDR4 RAM's one,
making it very convenient to fit all the data in the MCDRAM and avoid the
use of DDR4 for active data handling. The KNL offers 3 different
memory modes (see figure \ref{KNL4}):
\begin{itemize}
\item Flat Mode: In this mode the MCDRAM is treated as a NUMA node and
  the user controls what goes in MCRAM.
\item Cache Mode: In this mode the MCDRAM is treated as Last Level
  Cache and it is used automatically.
\item Hybrid Mode: This mode is a combination of the previous two,
  with the 
  ratio between Flat and Cache that can be chosen at boot time from the BIOS.
\end{itemize}
\end{itemize}
As reference the theoretical peak performance of the KNL using single
precision is of over 6 TFlop/s and over 3 TFlop/s for double precision
\cite{colfax}.

\subsubsection{Contributions}\label{sec-2.1}

\begin{figure}[thb] 
  \centering
 \hfill
   \subfigure[Wilson operator.]{\includegraphics[width=.4\textwidth]{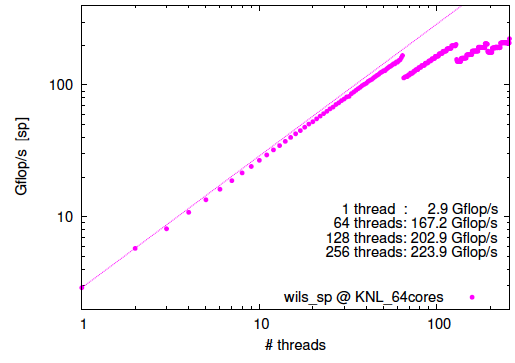}}
  \hfill
   \subfigure[Brillouin operator.]{\includegraphics[width=.4\textwidth]{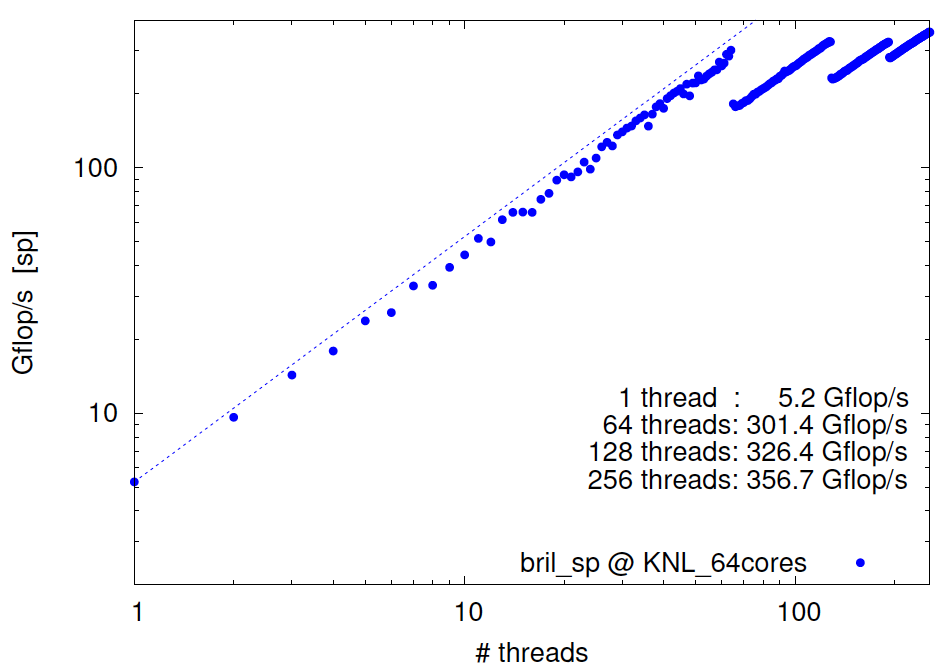}}
  \hfill
  \caption{Scaling of the flop count per second for the matrix vector
    operation as function of the number of threads.}
  \label{durr}
\end{figure}

During the conference many of the talks reported on
implementation and optimisation specific for the KNL architecture,
with a spectrum of coding philosophy ranging from the need of
portability and compatibility of the code, to implementations
specialised for the KNL architecture. \\[.3cm]
{\bf Optimisation of the Brillouin operator on the KNL architecture}:
S. Durr presented a rather straightforward, but still performing,
implementation of the Brillouin operator for the KNL architecture \cite{Durr:2017clx}.
Brillouin is an implementation of the Dirac operator that
shows better scaling properties of the dispersion relations
compared to the usual Wilson Dirac operator, and it represents a
suitable kernel for the overlap procedure \cite{Durr:2016xoc,Durr:2017wfi}.
Compared to the Wilson operator the Brillouin operator requires many
more hop terms, leading to an estimate of the ratio of computational
intensity of Brillouin over Wilson of 5/2.
The implementation strategy for the application of the operator on multiple
spinors can be schematically described as (from the innermost cycle to
the outermost cycle):
\begin{enumerate}
\item Unroll the local color and spinor indices.
\item SIMD-ize over the index of spinor with an OMP pragma.
\item Distribute the geometry of OMP threads.
\end{enumerate}
In figure \ref{durr} it is reported the flop count per second of the
single precision 
Brillouin and Wilson operators as function of the number of threads,
for number of colours $N_C=3$, number of spinor $N_V=3N_C$.
In the best configuration the Brillouin operator can obtain a performance
of $357/272$ Gflop/s (peak/sustained) meaning $6.8/10.4\%$ of the KNL
theoretical performances, while for the Wilson operator the
performances are $225/135$ Gflop/s  (peak/sustained) meaning  $4.3/
5.2\%$ of the KNL.
 
\begin{figure}[thb] 
  \centering
 \hfill
   \subfigure[Over-decompose and interleave elements from different virtual nodes in adjacent SIMD lanes.]{\includegraphics[width=.4\textwidth]{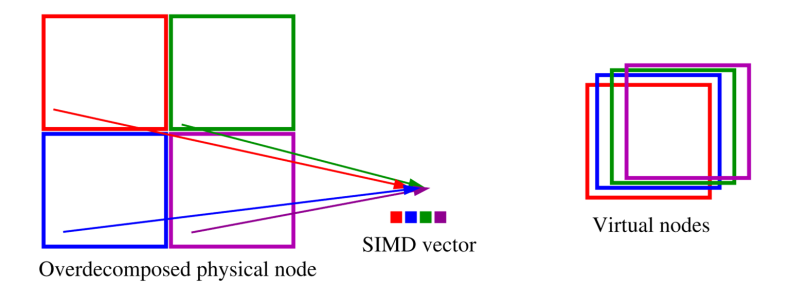}}
  \hfill
   \subfigure[SIMD accelerating matrix-vector products: Many vectors = many matrices x many vectors]{\includegraphics[width=.4\textwidth]{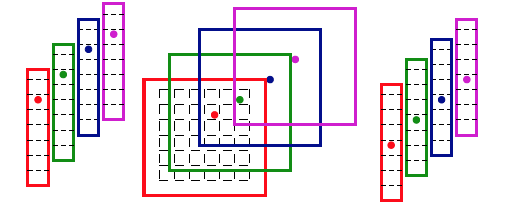}}
  \hfill
  \caption{Grid data layout \cite{Boyle:2015tjk}.}
  \label{grid1}
\end{figure}

The takeway information is that for this kind of applications a
reasonable  performance can be obtained, while maintaining full
portability even using simple coding strategy.\\[.3cm]
{\bf Grid software status and performance}: 
The data parallel framework Grid is developed by the
Edinburgh group: P. A. Boyle, G. Cossu, A. Yamaguchi and A. Portelli
\cite{Boyle:2015tjk,gitgrid}. One of the main features of the Grid
suite is the arrangement of the data layout, that can be summarised as:
\begin{itemize}
\item Geometrically decompose cartesian arrays across nodes (MPI).
\item Subdivide node volume into smaller virtual nodes.
\item Spread virtual nodes across SIMD lanes.
\item Use OpenMP+MPI+SIMD to process conformable array operations.
\item Automatically modify data layout to align data parallel operations to SIMD hardware.
\end{itemize}
The key idea is to execute exactly the same instructions on many
nodes, with each node acting at the same time on multiple virtual
nodes, see figure \ref{grid1}.  
The benefit comes from the observation that conformable array operations are simple and vectorise perfectly.
The performances of Grid in single precision on single node for multiple
RHS on the Wilson operator were already presented 
\cite{Boyle:2015tjk,Boyle:2016lbp,Boyle:2017vhi}, it is worth however to highlight the most
relevant features:
\begin{itemize}
\item Single core instructions-per-cycle (IPC) is 1.7  ($85\%$ of theoretical, 2 IPC).
\item Multi-core L1 hit rate is $99\%$ (perfect SFW prefetching).
\item Multi-core MCDRAM bandwidth $97\%$ (370GB/s), 
with 1 thread per core fastest after writing in assembler (not intrinsics).
\end{itemize}
The group reports a very detailed breakdown of the data movement
trough the system caches and compare their findings with the
optimal theoretical values, see table \ref{tab-grid}.
\begin{table}[thb]
  \small
  \centering
  \caption{Grid data movement.}
  \label{tab-grid}
  \begin{tabular}{ccccc}\toprule
L1 read	&L1 write&	L2 read	&MCDRAM read&	MCDRAM write \\\midrule
550	&12 (12 theo.)&	98 (96 theo.)	&68 (12 theo.)& 	12 (12 theo.) \\\bottomrule
  \end{tabular}
\end{table}

From the ratio of the L2 read over the MCRDAM read they evaluate the
reuse factor ($98/68=1.44$) that is well below the ideal case
of infinite L2 capacity ($96/12=8$). The data seems to suggest this
inefficiency is due to a non optimal scaling of the interconnection mesh of
the L2 cache on the KNL. 
Finally for the single node single precision assembler Domainwall Dirac operator and with the
described data layout Grid achieves a performance on the  KNL 7250 of 960
Gflops/s meaning a $16\%$ of the peak performance. The multi node
performances are discussed in section \ref{sec-4.1}.\\[.3cm]
{\bf Wilson and Domainwall Kernels on Oakforest-PACS}:
I. Kanamori and H. Matsufuru developed for Bridge++ two different
implementations of a Wilson/Domainwall kernel
\cite{Kanamori:2017tlp}. 
 Their strategy was to have a direct comparison of two radically
different implementations: a simple one (impl-1) and a more aggressive
one (impl-2). The most notable features of the two implementations are
summarised in table \ref{tab-oak}.
\begin{table}[thb]
  \small
  \centering
  \caption{Features of the two kernels implementation.}
  \label{tab-oak}
   \begin{tabular}{l|l}
\toprule
Simple (impl-1) see figure \ref{oak}.a                                                    &     Aggressive (impl-2) Grid-like    see figure \ref{oak}.b                        \\\midrule                                
Simd vector is continuously packed in x-direction.                 &     Simd vector is distributed to subdomains        \\            
No MPI parallelization in x- direction.                            &     Non-blocking communications.                                              \\
Blocking communications.                                           &     (partial) Loop tiling.                                                    \\
No manual prefetch.                                                &     Manual prefetch.                                                          \\
AVX512 intrinsics for arithmetics.                                 &\\\bottomrule
  \end{tabular}
\end{table}

\begin{figure}[thb] 
  \centering
 \hfill
   \subfigure[Implementation 1 (Simple)]{\includegraphics[width=.4\textwidth]{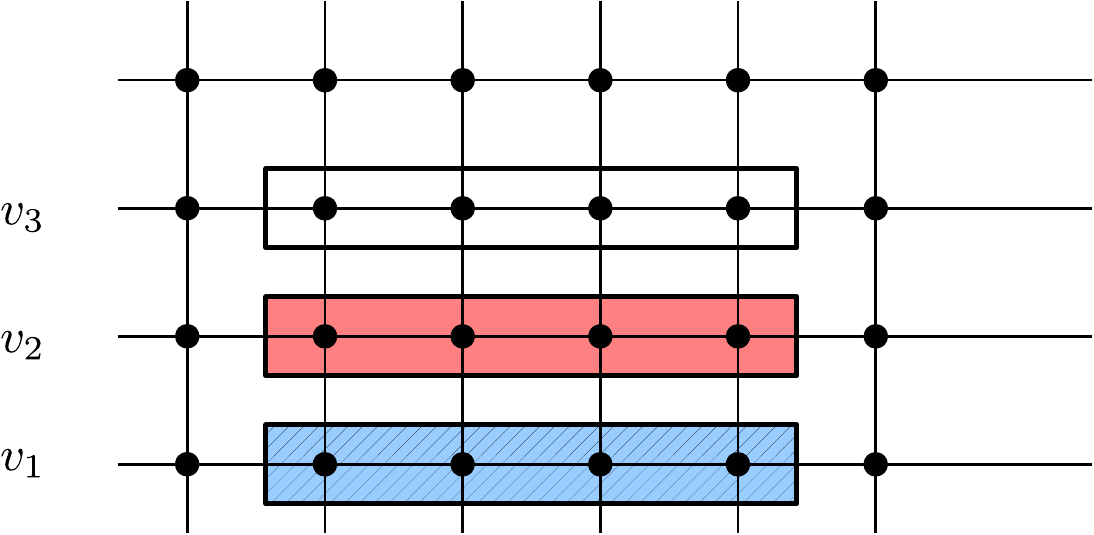}}
  \hfill
   \subfigure[Implementation 2 (Aggressive)]{\includegraphics[width=.4\textwidth]{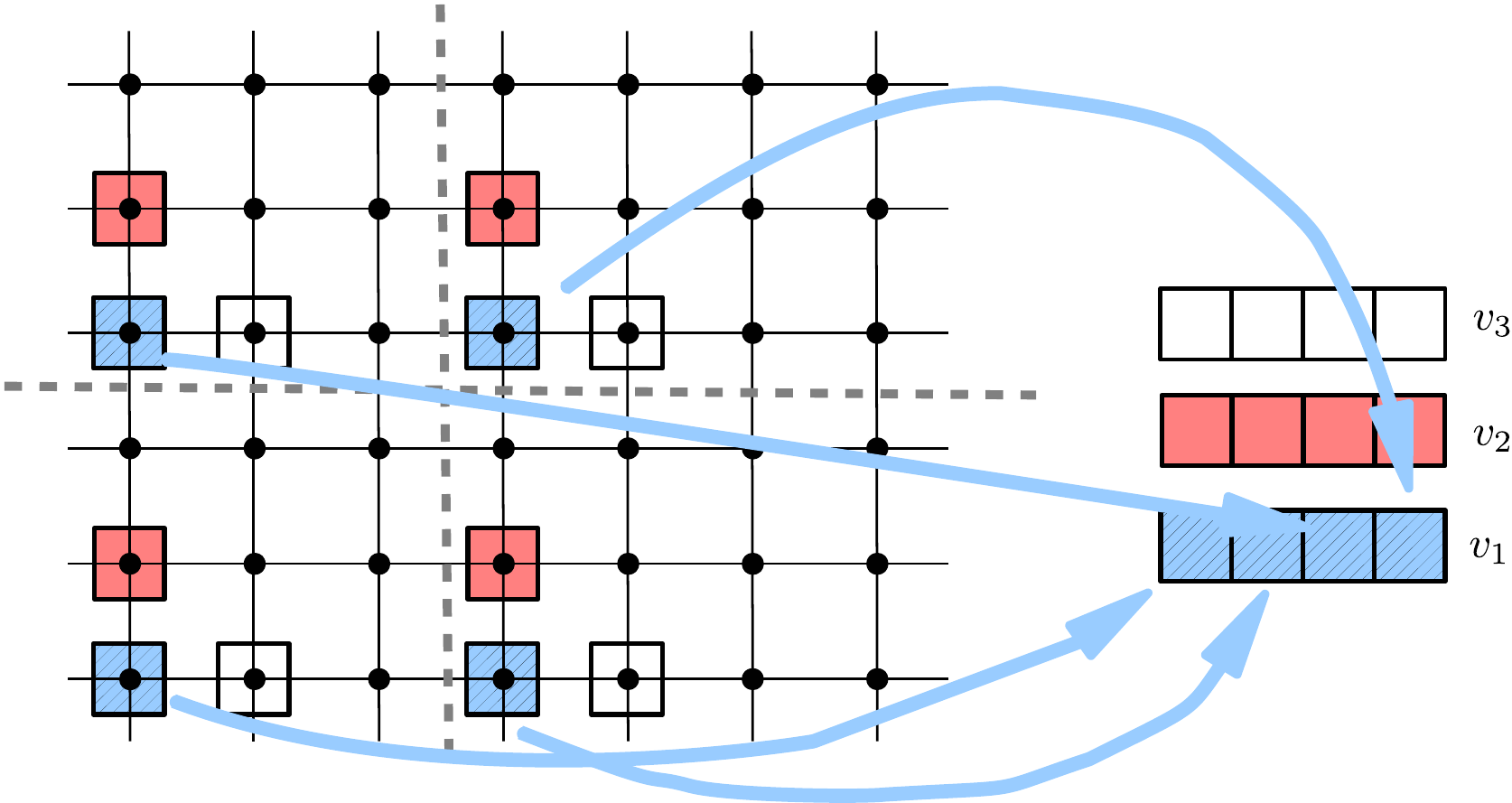}}
  \hfill
  \caption{Data layout for two implementations of the Wilson/Domainwall operator.}
  \label{oak}
\end{figure}
The authors reported both on the single node and on the scaling performances of their codes on the
OakForest-PACS PRIMERGY CX1640 M1 by Fujitsu (Intel Xeon Phi 7250 68C
1.4GHz (KNL) + Intel Omni-Path network topology: Full-bisection Fat
Tree).

For the single node performances of the Wilson Dirac operators on a  $32^3\times 64$ lattice results
are:
\begin{itemize}
\item  Impl-1: 241 GFlops (single precision, 4 threads/core: totally 256 threads), 147 GFlops (double, 4 threads/core)
\item  Impl-2: 339 GFlops (single, 2 threads/core), 174 GFlops (double, 2 threads/core)
\end{itemize}

\begin{figure}[hbt] 
  \centering
 \hfill
   \subfigure[Weak scaling]{\includegraphics[width=.4\textwidth]{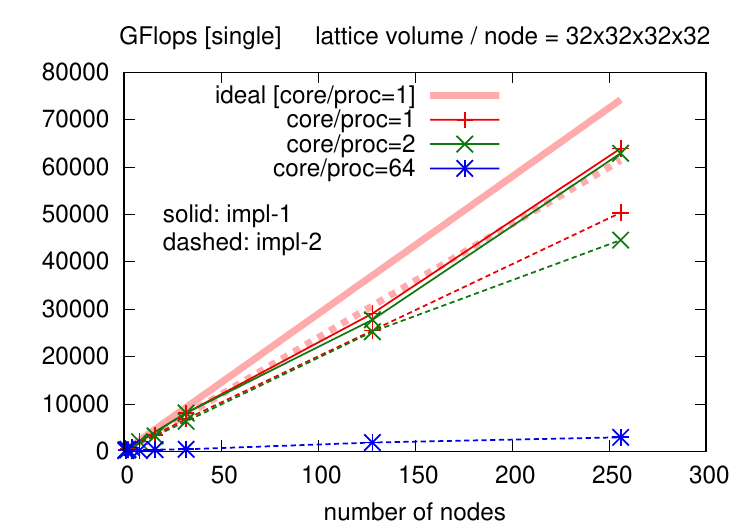}}
  \hfill
   \subfigure[Strong scaling]{\includegraphics[width=.4\textwidth]{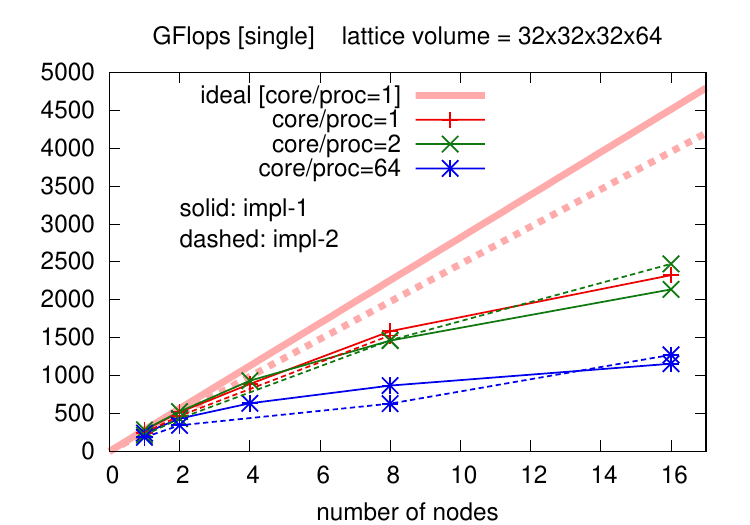}}
  \hfill
  \caption{Scaling plots of the Wilson operator for impl-1 (solid) and
    impl-2(dashed). Ideal scalings are plotted with pink thick lines.}
  \label{oak2}
\end{figure}
They noted a strong dependence on the number of the
threads/core for impl-1, with an increase of performance when using
larger number of threads/core, while such behaviour is only marginally
present in impl-2.

About the scaling performances the authors presented their studies for
weak and strong scaling, as reported in figure \ref{oak2}.

They also reported on the single node and scaling performances of the
domain wall operator with the impl-2 largely outperforming impl-1.
The best performance for the single node study were obtained with 32 MPI processes and 4
threads/core: 395 GFlops for the single precision case and 197
GFlops for the double precision.

 As a general tendency they noted that both for the Wilson and for the
 domainwall operators, the 1 and 2 cores/process cases give better performance than the 64
 cores/process case. \\[.3cm]
{\bf MILC code performance on high end CPU and GPU supercomputer clusters}:
R. Li reported on the performances of three of the major  QCD routines of the
MILC code: the staggered multi-shift CG, Symanzik one-loop
gauge-force, and the HISQ fermion force on the KNL architecture\cite{Li,milc}.
For the staggered multishift CG and the Symanzik force the authors compared the performances obtained
with the use of two different set of libraries: QOPQDP, the MILC
baseline and QPhiX, the KNL optimised library developed in
collaboration with Intel. 
The QPhiX library features the use a structure-of-array (SOA) data structure for improved
cache reuse and a data layout similar to the Grid implementation.

The authors report an increase in perfomance of a factor 1.5 on the CG
in using QPhiX on one KNL, while for the multinode application no
substantial increase was found signalling that the algorithm was
network bound on their test machine. 

\begin{figure}[thb] 
  \centering
 \hfill
   \subfigure[CG speedup.]{\includegraphics[width=.4\textwidth]{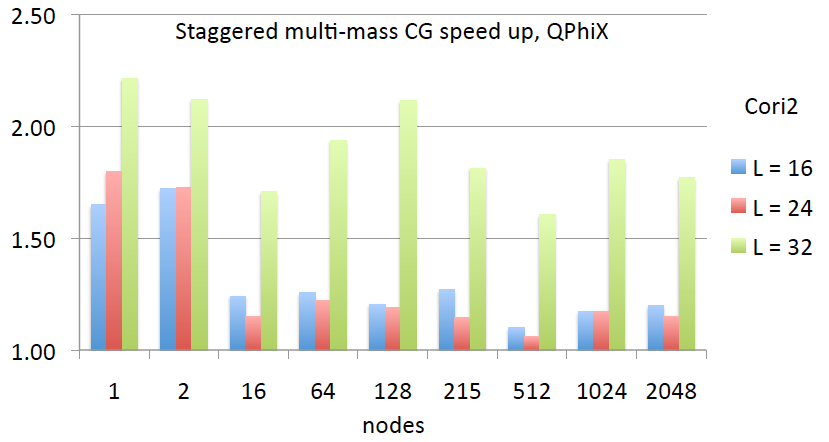}}
  \hfill
   \subfigure[Symanzik gauge speedup.]{\includegraphics[width=.4\textwidth]{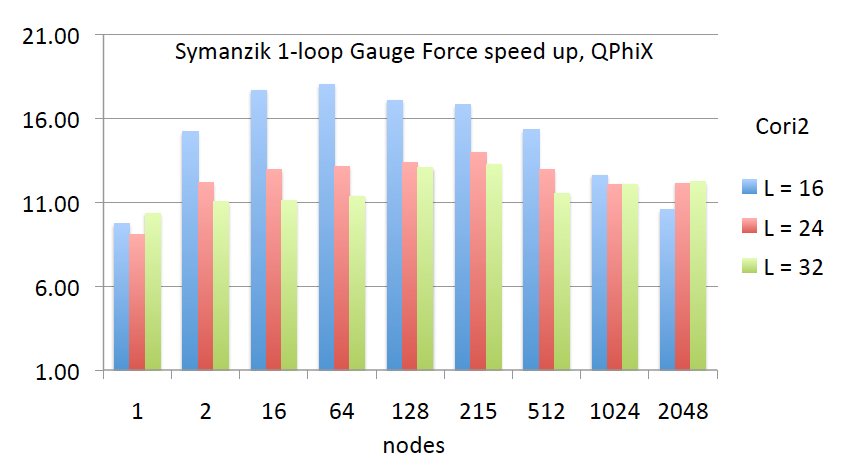}}
  \hfill
  \caption{QPhiX library speedup over the MILC baseline on NERSC - Cori II}
  \label{MILC}
\end{figure}

They also report on the weak scaling performances of the gauge and
fermion forces with a scaling efficiency as high as $80\%$ (see table
\ref{tab-milc} and figure \ref{MILC}).

\begin{table}[thb]
  \small
  \centering
  \caption{NERSC - Cori II single node and scaling performances:
    comparison among the MILC baseline and the QPhiX libraries.}
  \label{tab-milc}
  \begin{tabular}{c|cc|cc}\toprule
 &\multicolumn{2}{|c|}{multi-mass CG} &\multicolumn{2}{|c}{Symanzik Gauge force}\\\midrule
&	MILC b.&	QPhiX  &	MILC b.&	QPhiX  \\
Gflops, single node	&70&	120&	60 &	380 (kernel)\\
Scaling factor $(16\leq L\leq 32)$&0.3$\sim$0.55&0.25$\sim$0.45&0.4$\sim$0.85&0.65$\sim$0.9\\\bottomrule
  \end{tabular}
\end{table}

\subsection{Tesla Volta}
\begin{figure}[thb] 
  \centering
 \hfill
   \subfigure[GV100 structure.]{\includegraphics[width=.3\textwidth]{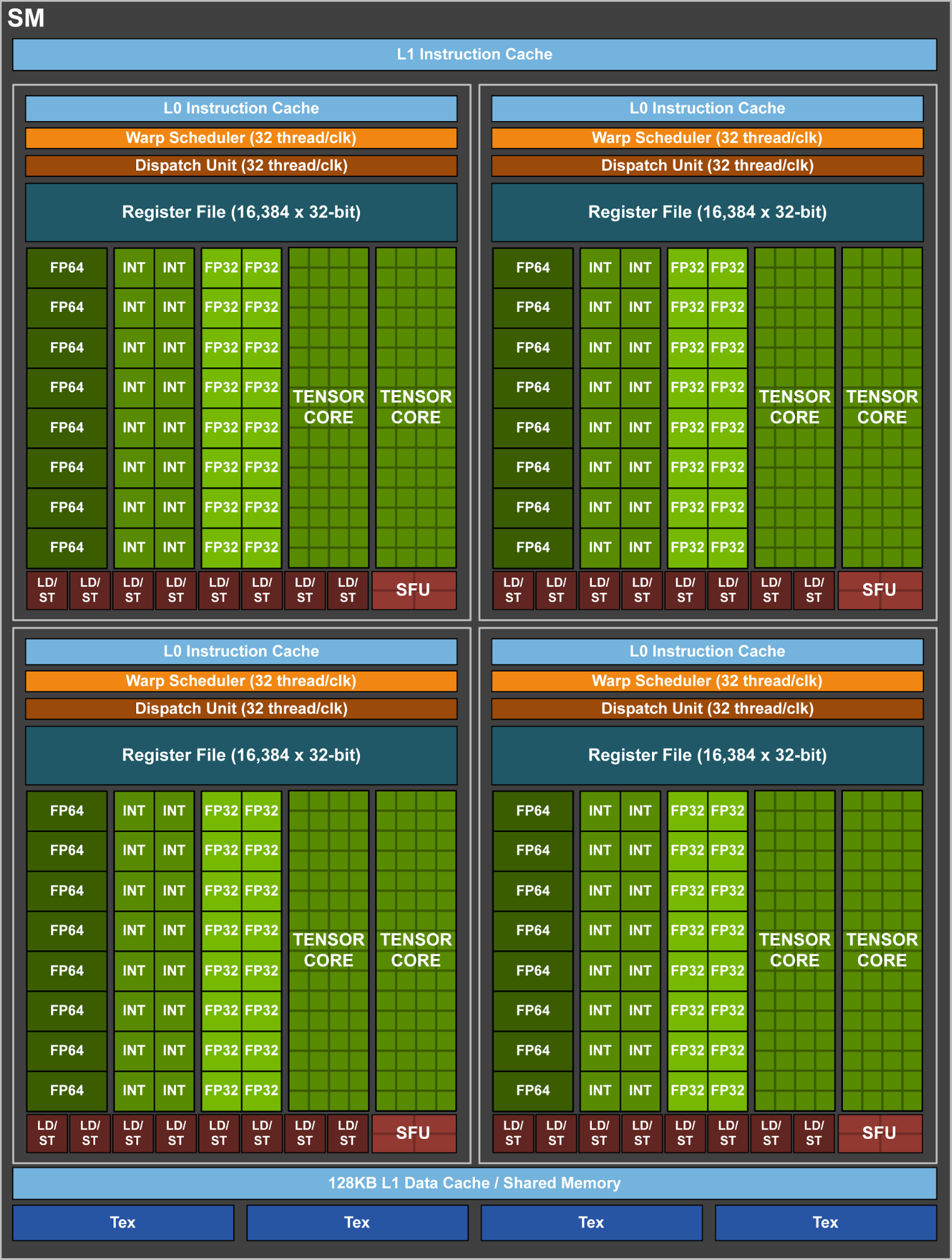}}
\hfill
   \subfigure[Volta vs. Pascal.]{\includegraphics[width=.4\textwidth]{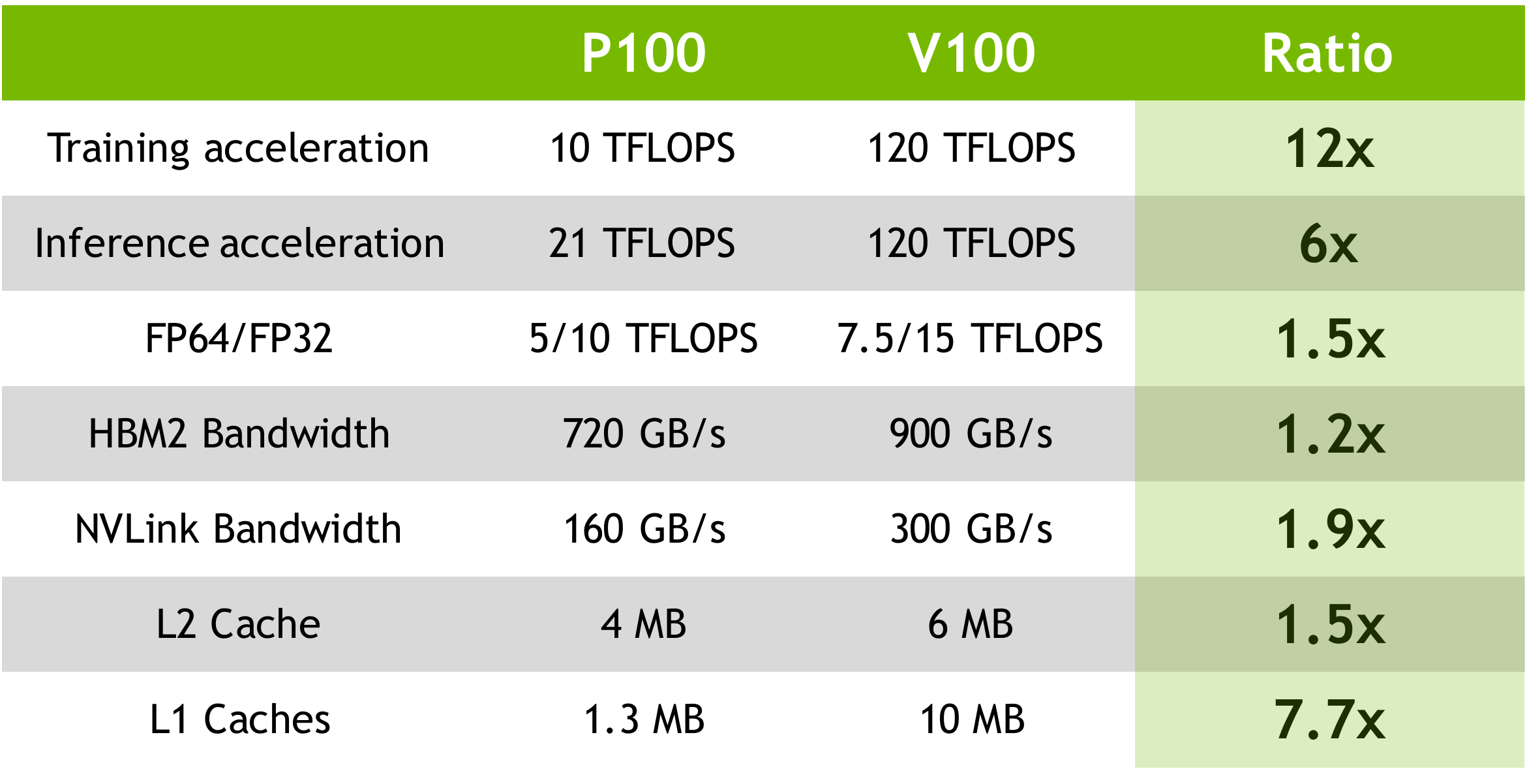}}
\hfill
  \caption{NVIDIA Volta technical details.}
  \label{volta}
\end{figure}

Tesla V100 is latest NVIDIA accelerator based on the new NVIDIA
Volta GV100 GPU \cite{volta}.
The GV100 GPU includes 21.1 billion transistors with a size of 815
mm2.
In figure \ref{volta}.a there is a schematic breakdown of the GV100
Chip. A full GV100 GPU consists of six Graphics Processing Clusters
(GPCs), 84 Volta Streaming Multiprocessors (SMs), 42 Texture
Processing Clusters (TPCs) each including two SMs, and eight 512-bit memory controllers (4096
bits total). 

The Tesla Volta features the second generation NVLink, that allows for both GPU-GPU and GPU-CPU
interconnections. In particular the GV100 supports up to 6
NVLink links with 25 GB/s bandwith for a total aggregate bandwidth of 300 GB/s. 
In figure \ref{volta}.b is reported a direct comparison of the
hardware features with the previous GPU generation (Pascal).

To summarise the main improvements for lattice QCD come from:
\begin{itemize}
\item 1.5x faster sustained memory bandwidth.\\
Volta sustains $95\%$ of peak memory bandwidth in STREAM.
\item Faster NVLink for better multi-GPU performance (300GB/S).
\item Improved L1 Data Cache and Shared Memory subsystem.\\
 The  combined capacity is 128 KB/SM, more than 7 times larger than the
  previous generation (GP100) data cache.
\end{itemize}

\subsubsection{Contributions}\label{sec-3.1}

\begin{figure}[thb] 
  \centering
\hfill 
   \subfigure[Single precision Wilson-Dslash operator performances for different hardware]{\includegraphics[width=.44\textwidth]{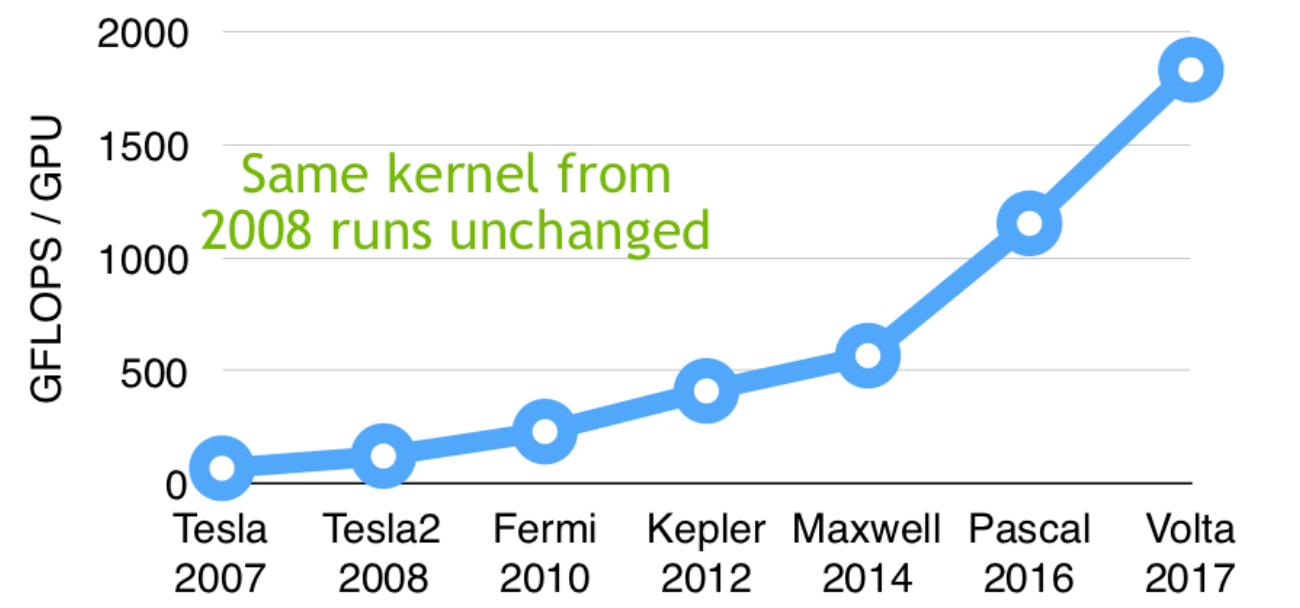}}
\hfill 
   \subfigure[Hisq Dslash for multiple rhs]{\includegraphics[width=.44\textwidth]{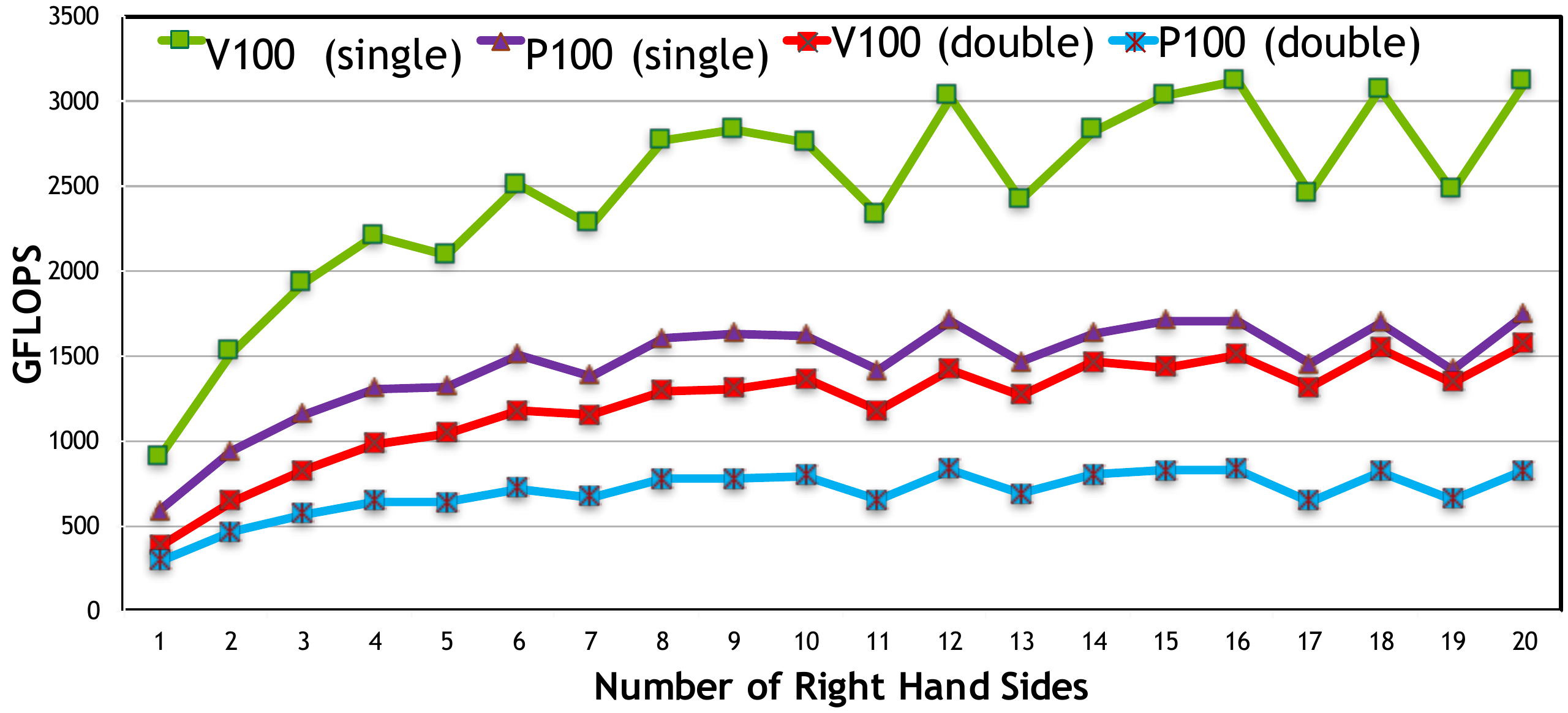}}
\hfill 
  \caption{QUDA single GPU performances}
  \label{quda1}
\end{figure}

{\bf Developing QCD Algorithms For NVIDIA GPUs Using the QUDA
  Framework}: 
QUDA is an NVIDIA GPU library for QCD simulations, it can be used both to
accelerate traditional lattice QCD applications or as standalone framework.
It supports many different fermion discretizations, and features many
algorithms like Adaptive Multigrid, deflation and Block Krylov space
methods, using various techniques such as mixed-precision and
communication hiding to maximise performance \cite{Clark:2009wm,Babich:2011np,quda}. This library has been
around for roughly ten years and has kept on showing increased
performances as the hardware evolved with time, showing very good
properties of portability across the NVIDIA GPU generations (see figure \ref{quda1}.a). In particular for
the application of the same Hisq Dslash over multiple righthand sides the
scale factor from Pascal to Volta is roughly a factor two  (see figure
\ref{quda1}.b).

\begin{figure}[thb] 
  \centering
\hfill 
   \subfigure[Node structure.]{\includegraphics[width=.3\textwidth]{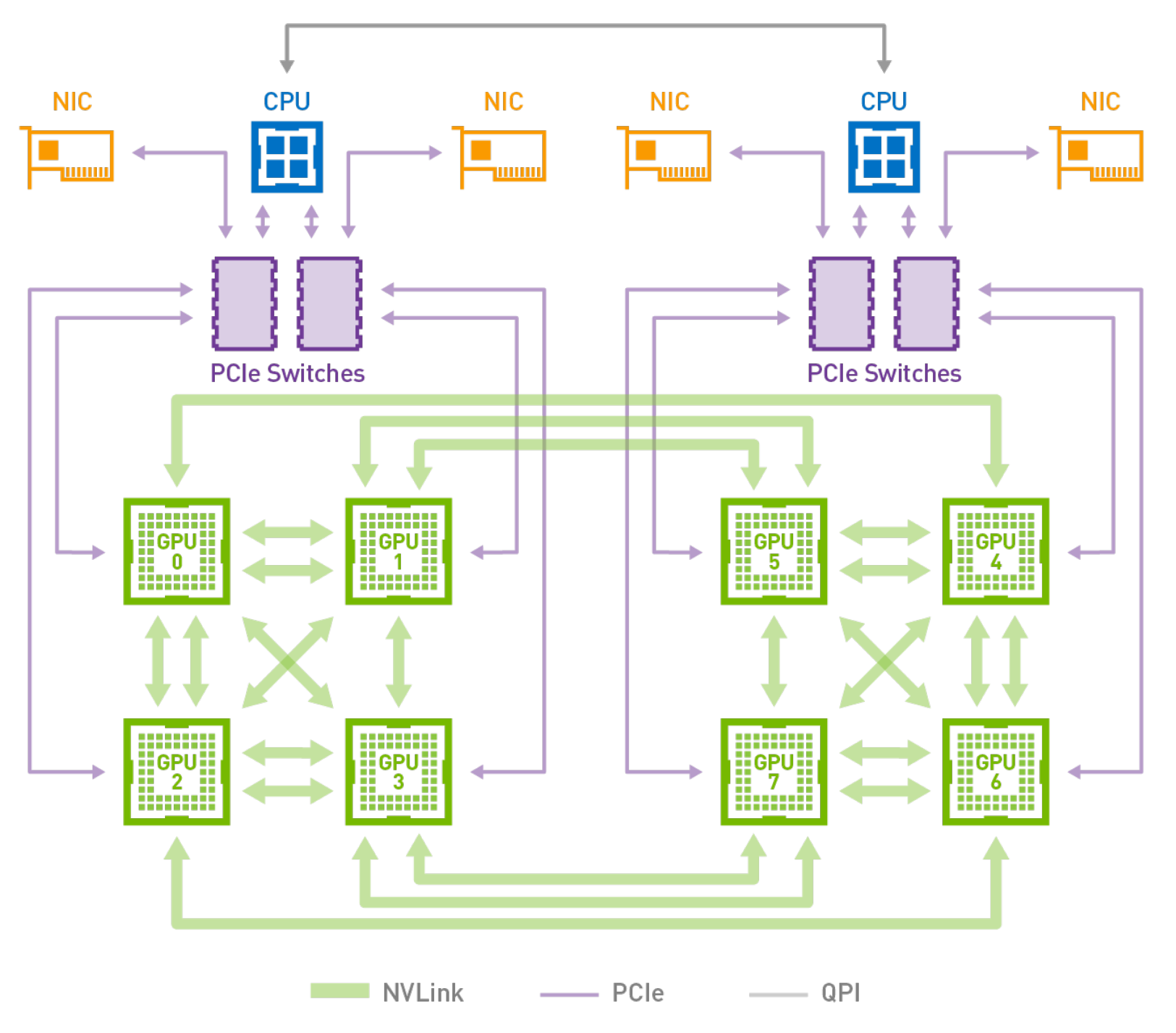}}
\hfill 
   \subfigure[Solver (mixed precision Shamir) weak scaling.]{\includegraphics[width=.44\textwidth]{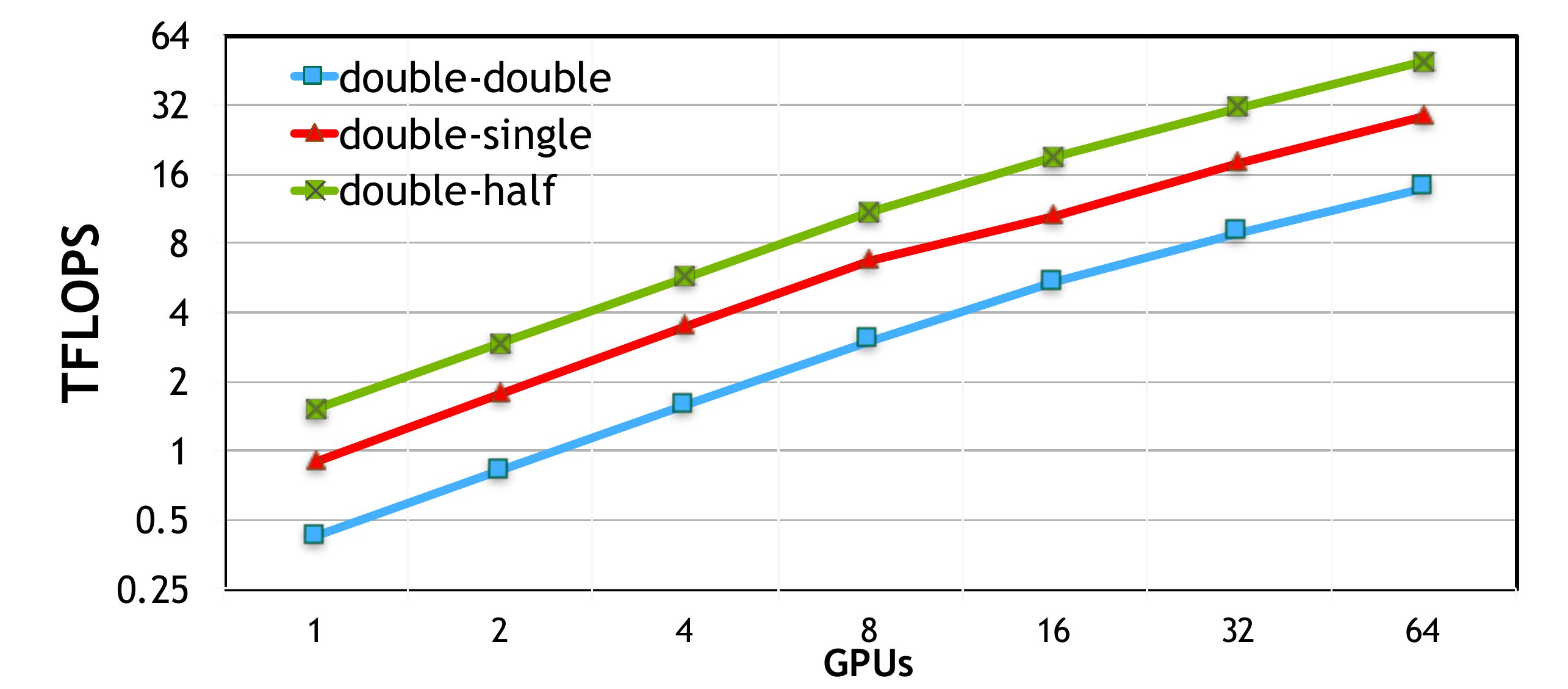}}\hfill 
\\
\hfill 
   \subfigure[Clover strong scaling.]{\includegraphics[width=.44\textwidth]{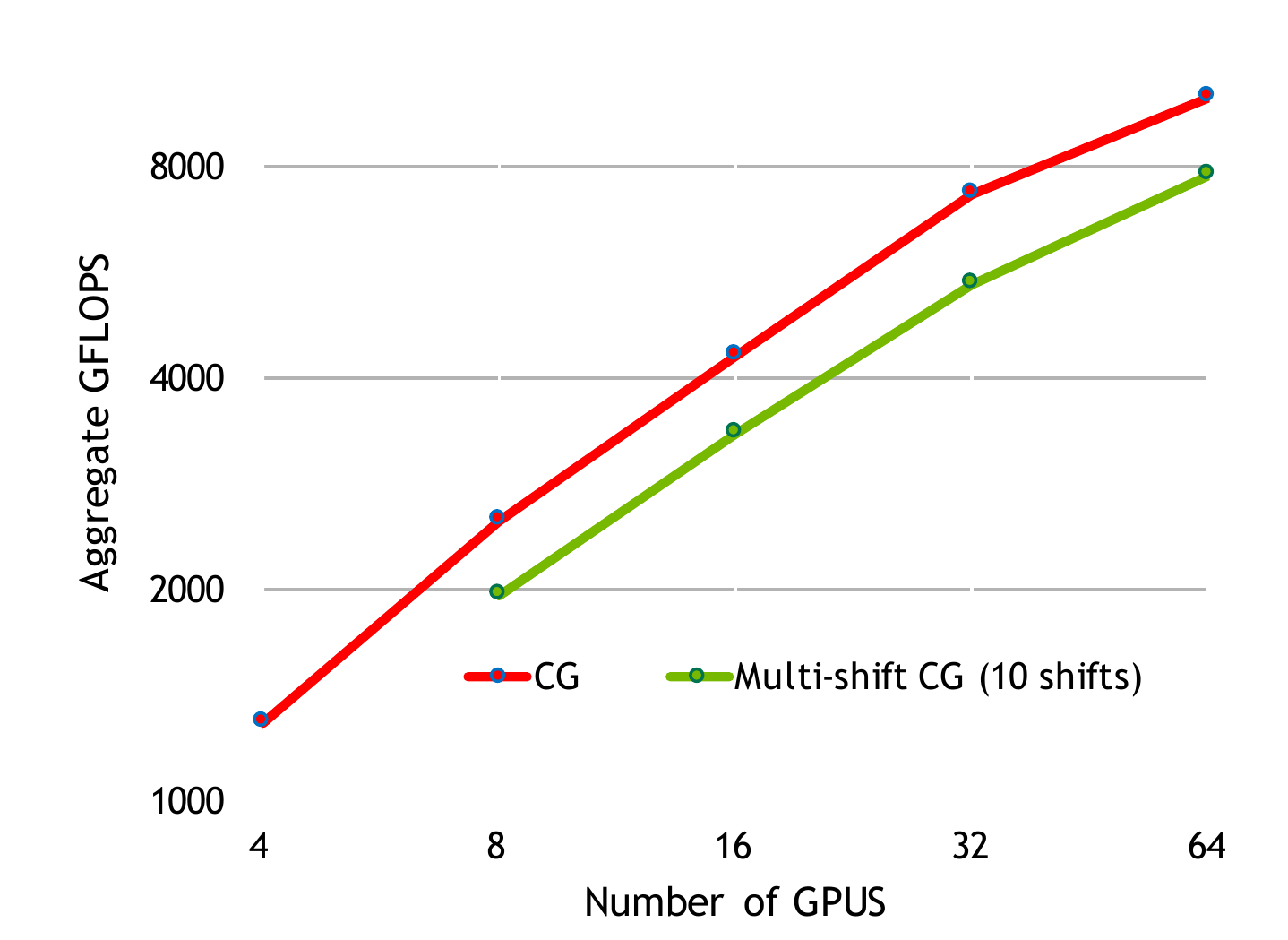}}
 \hfill 
  \subfigure[HISQ strong scaling.]{\includegraphics[width=.44\textwidth]{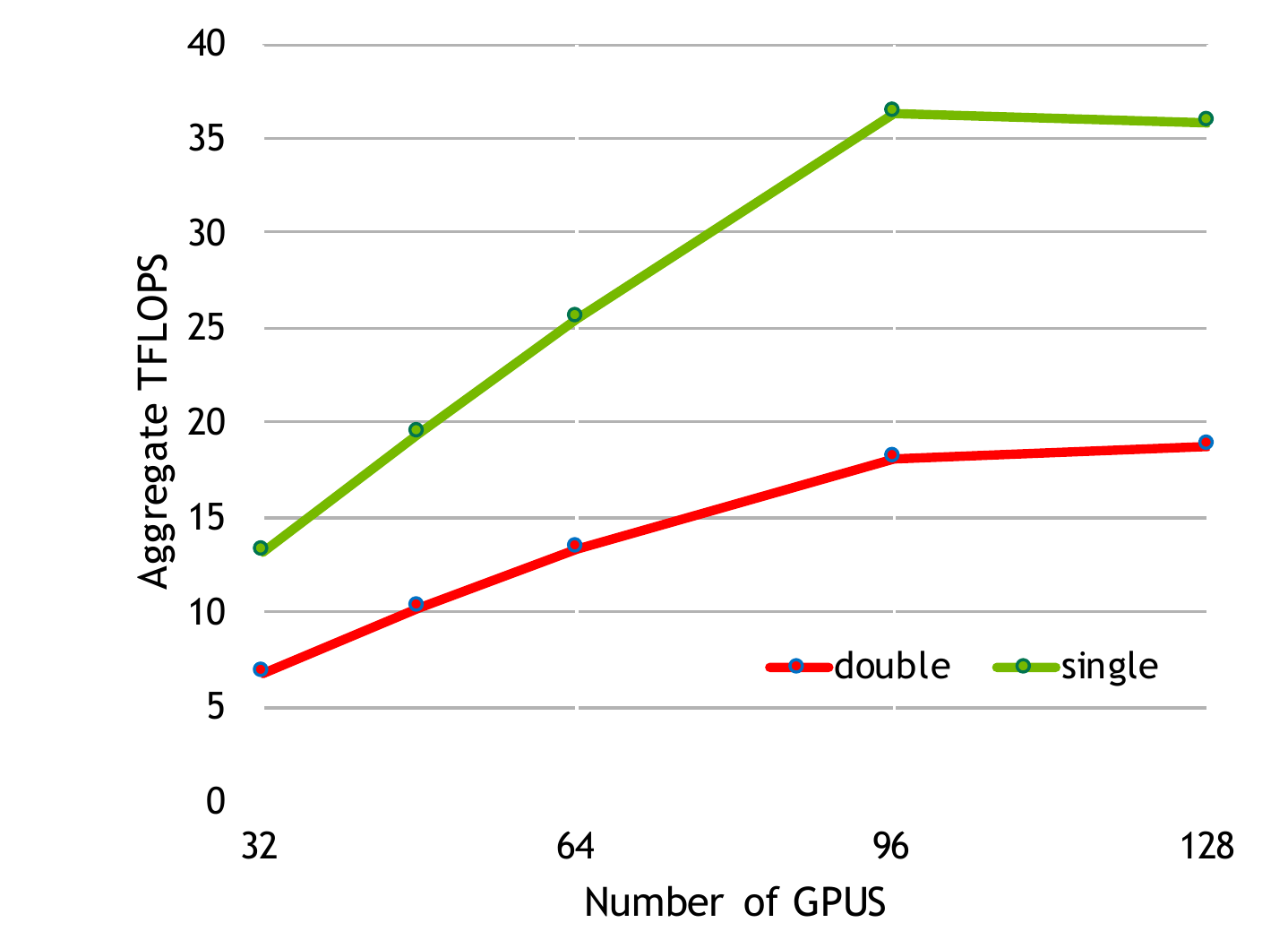}}\hfill 
\\
  \caption{QUDA multi-GPU scaling performances.}
  \label{quda2}
\end{figure}

K. Clark and M. Wagner \cite{clark,wagner} presented the scaling performances of the QUDA
libraries both for weak and strong scaling on a cluster equipped with
8xP100 (Pascal) GPUs nodes, with NVLINK as intra-node GPU connection and using 4x EDR
for inter-node communication (see figure \ref{quda2}.a).

The code shows good scaling properties, but what is particularly
instructive is the breakdown of the steps needed to achieve such
scaling behaviour: 
\begin{itemize}
\item For the intra-node, direct peer-to-peer
communication.
\item Direct access to the network interface through GPU Direct for
  inter-node communication.
\item Topology aware communications, in order to avoid multiple hops
  among the PCIe switches.
\end{itemize}
{\bf Performance Portability Strategies for Grid C++ Expression Templates}:

M. Lin presented successes and challenges encountered in porting the Grid C++ expression
template to GPU-based systems and exploring extensively different
approaches to integrate CUDA, OpenACC and Just-In-Time compilation \cite{Boyle:2017gzg}. 

In order to port the Grid library in an “effortless” way the authors
strategy was to map the Grid's block into a vector and the threads into
the elements of a vector. This approach greatly simplifies
code integration, because most of the underlying code structures are
shared.

\begin{figure}[thb] 
  \centering
\hfill 
   \subfigure[Performance over several architectures. The vertical
   lines indicate the cache size, the horizontal ones show the memory
   bandwidth obtained in a stream triad test.]{\includegraphics[width=.45\textwidth]{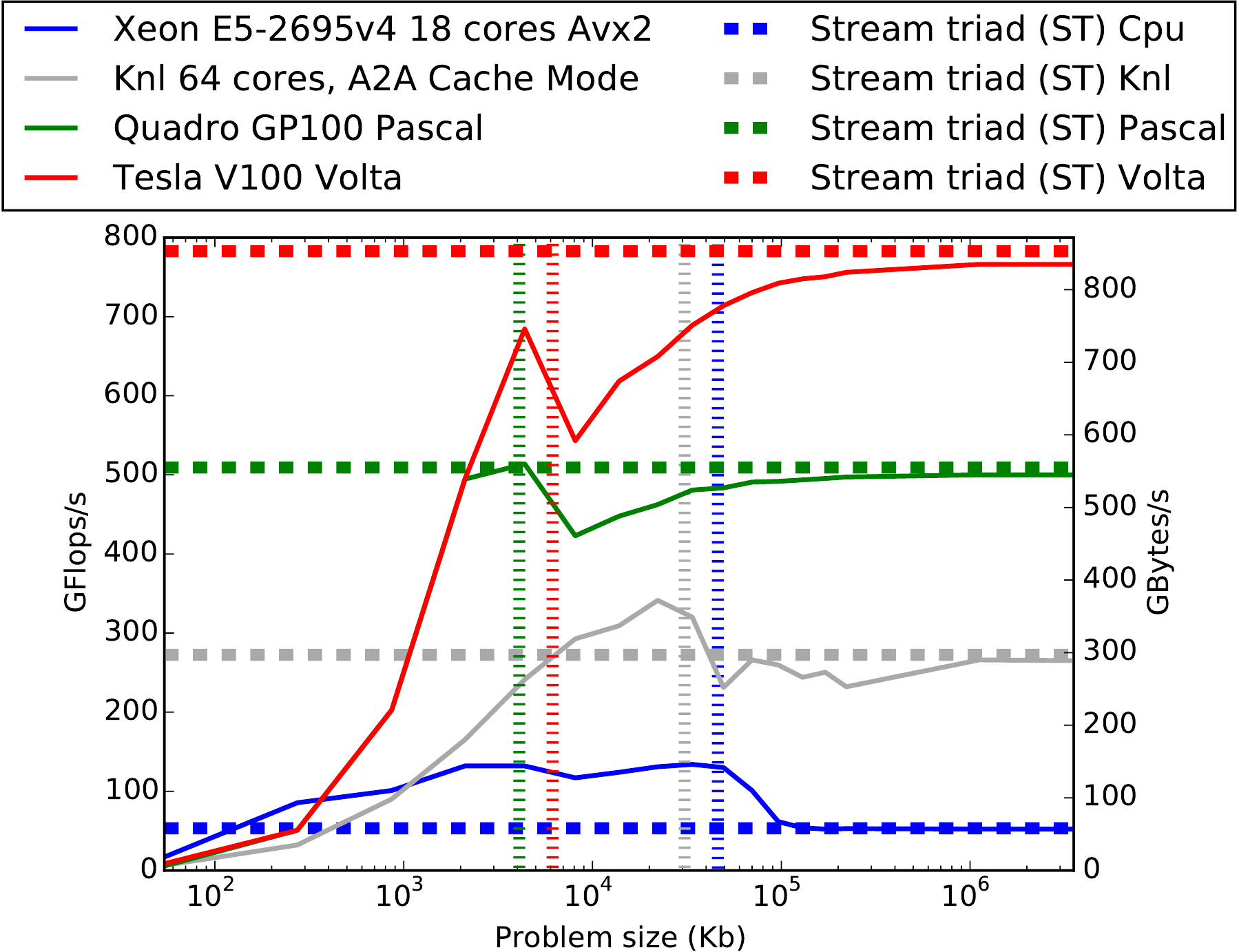}}
 \hfill 
   \subfigure[Performance as a function of the vector length in a Quadro GP100. ]{\includegraphics[width=.45\textwidth]{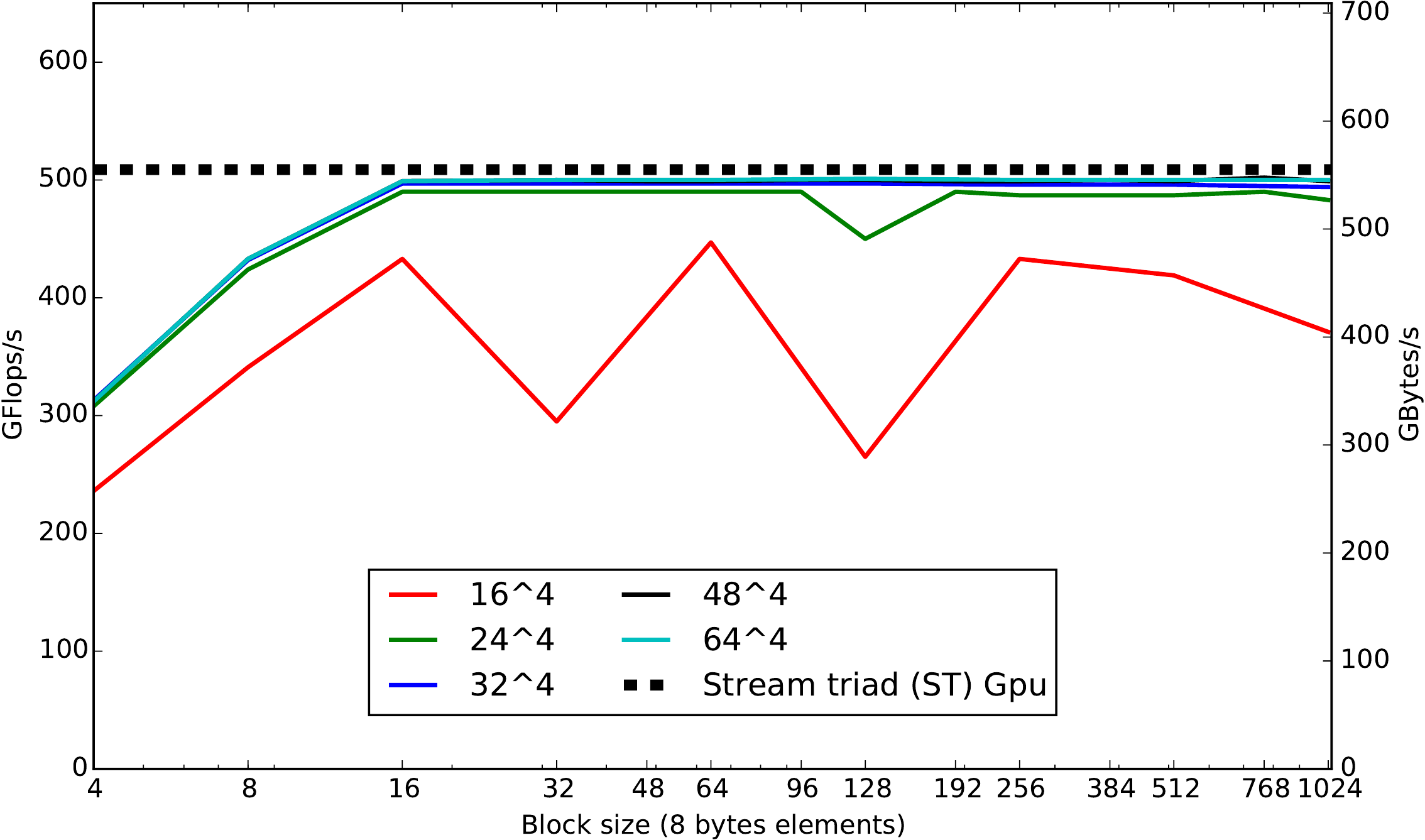}}
 \hfill \caption{SU(3)$\times$SU(3) streaming test over multiple architectures.}
  \label{gridtemplate}

\end{figure}

They report on the performance studies in the case of an SU(3)$\times$SU(3)
streaming test, showing that their approach is capable of achieving good
performance (see figure \ref{gridtemplate}).

In particular in figure \ref{gridtemplate}.b  the authors show that
the performance reaches its maximum with a block size of 16 threads, or
equivalently a vector length of 16 single-precision complex numbers. 

They also comment on possible improvements to avoid the need of large
vector length.

The porting of the Grid Dslash operator is under
active development.
 
\subsection{Intel Omni-Path}\label{OPA}
Omni-Path (OPA) is the new  high-performance communication architecture
designed by Intel. It is designed specifically for HPC purpose with
low communication latency, low power consumption and a high
throughput. 
It is developed by Intel as the on-ramp to exascale computing.

 The OPA is designed and
 optimised for the MPI communication model and it is capable of a scalable All-To-All communication.

One of the principal differences with other comparable products is that
all the network functions are executed by the hosting system CPU as the
OPA doesn't equip any co-processor. As such overlapping communications and computations will anyway have to
rely on a common set of resources.

One of the key issues that was highlighted multiple times during the
conference was the unavailability of the threaded version of the
communication driver in the Omni-Path software stack. This limitation has been overcome in the recent major
upgrade \cite{psm2}.

In our community already quite a few production machines are equipped with
this new fast network and reports on their efficiency and scalability
were presented during the conference.

\subsubsection{Contributions}\label{sec-4.1}
{\bf An in-depth evaluation of the Intel Omni-Path network for LQCD
  applications} and {\bf An implementation of the DD-$\alpha$ AMG multigrid solver on Intel Knights Landing}:
P. Georg and D. Richtmann presented the porting of their collaboration codebase onto the new
QPACE3 machine, based on KNL nodes \cite{Georg:2017zua}. More
specifically they presented their
efforts in porting of the DD-$\alpha$AMG solver. QPACE 3 is
equipped with the new OPA interconnection and a very in-depth analysis
of the performances and issues associated with this new hardware was
presented.
The authors have highlighted as key performance
factors the network latency and the message rate. 
They noted that, when running on a single
or a small number of nodes, the standard parallelization approach with
one MPI rank per node and one or more threads on each core gives the
best performances on their code. 
On the contrary, when running on larger number of nodes, they found
beneficial using multiple MPI ranks per processor. 

\begin{figure}[thb] 
  \centering
\hfill 
   \subfigure[Bandwith as function of number of processes]{\includegraphics[width=.45\textwidth]{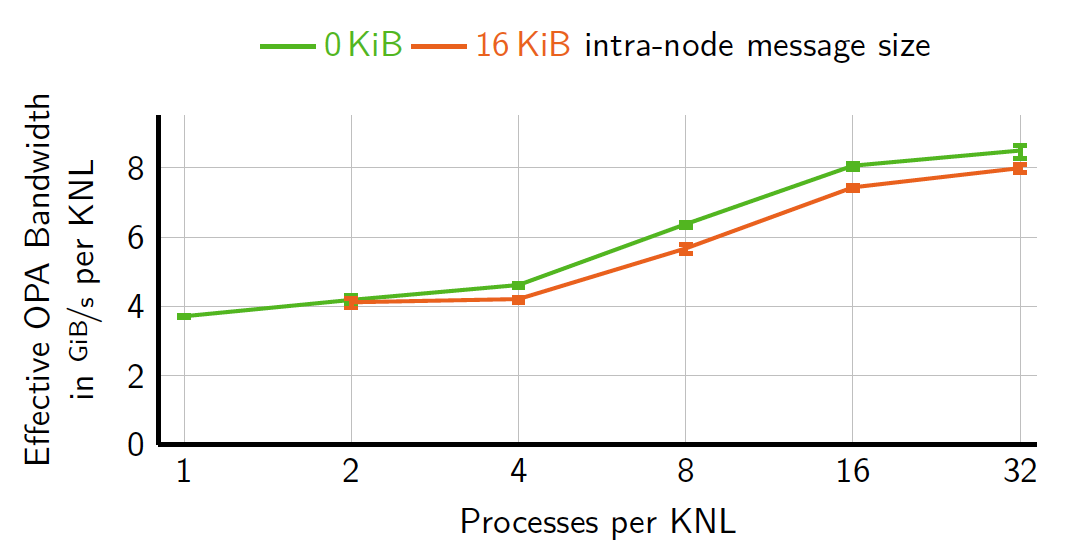}}
\hfill 
   \subfigure[Speedup as function of the number of processes]{\includegraphics[width=.45\textwidth]{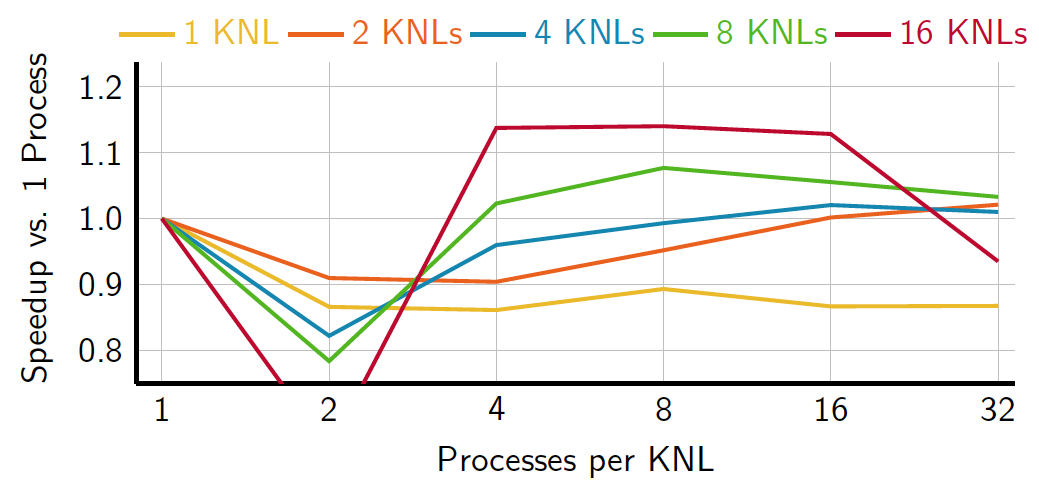}}
\hfill 
   \subfigure[DD-$\alpha$AMG strong scaling on KNL  and KNC]{\includegraphics[width=.45\textwidth]{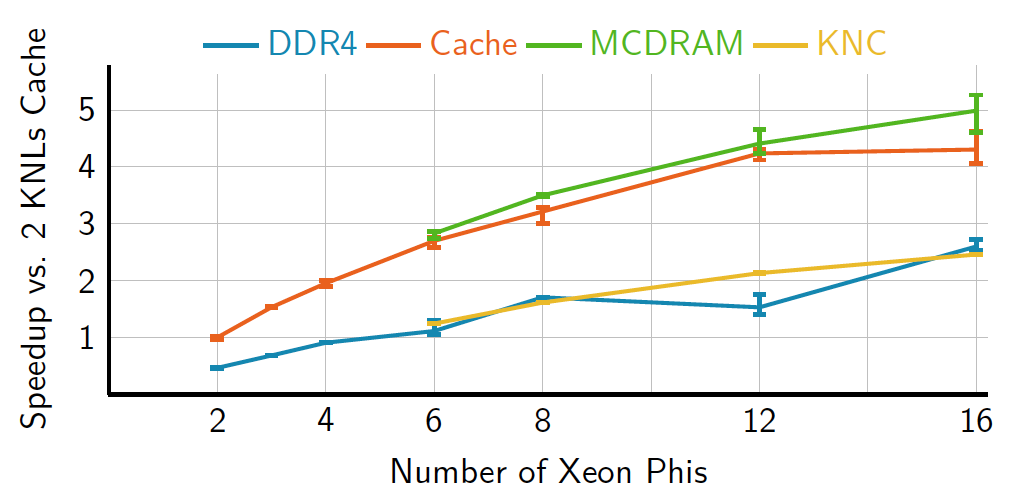}}
\caption{Performance analysis of the Intel OPA network.}
  \label{opa1}
\end{figure}

In particular in figure \ref{opa1}.a they presented the halo-exchange
performances on 16 KNL. They present the aggregate bandwidth
in GiB/s of a single KNL as function of the number of processes for
inter-node message size of 512 KiB / (number of processes per node)
and 0 and 16 KiB intra-node message size.

In figure \ref{opa1}.b it is reported the speedup factor of the DD-$\alpha$AMG for
various size of cluster of KNLs vs. number of processes per KNL. 

Finally in  figure \ref{opa1}.c  they present the comparison in speedup
between KNC and KNL for the off-chip strong scaling of DD-$\alpha$AMG.\\[.3cm]
{\bf Grid software status and performance}: The Grid developers have
analysed the performances of the OPA infrastructure and compared
them with the performances of other network hardware. They have studied the Grid's halo
exchange performances on three different machines
respectively equipped with Intel-OPA, Mellanox-EDR and Cray-Aries. They highlighted that a single MPI-process per
node on OPA is not capable of saturating the bandwith for the typical
message size used by Grid, leading to the need of subdividing the
problem within a node, hence increasing the unwanted intranode communication.
A short summary of their findings is reported on table \ref{tab-network-grid}.

\begin{figure}[thb] 
  \centering
\includegraphics[width=.65\textwidth]{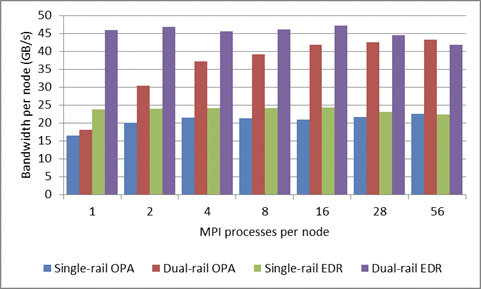}
\caption{Comparison of performances of the Grid's halo exchange between Intel OPA and Mellanox EDR dual rail  (courtesy: SGI HPE)}
  \label{network-grid}
\end{figure}

\begin{table}[thb]
  \small
  \centering
  \caption{Grid halo-exchange best performances on multiple architecture}
  \label{tab-network-grid}
  \begin{tabular}{llllll}\toprule
Machine & Node type & Network & Bidi peak GB/s & 1 rank per node GB/s& 4 ranks per node GB/s\\\midrule
BNL& KNL& Dual OPA& 50& 14& 44\\
Cori2& KNL& Aries& $\sim$16& 11 &11\\
SGI &ICE-X Xeon& Dual EPR &50 &44& 44\\\bottomrule
  \end{tabular}
\end{table}  
In figure \ref{network-grid} it is reported  a more direct comparison between Intel OPA and Mellanox EDR
dual rail on SGI ICE-X Xeon E5-2690 2.4Ghz nodes(only best performances).

\begin{figure}[thb] 
  \centering
\hfill 
   \subfigure[Geometrical mapping: different ranks on the same nodes
   belong to a cube of next-neighbours.]{\includegraphics[width=.30\textwidth]{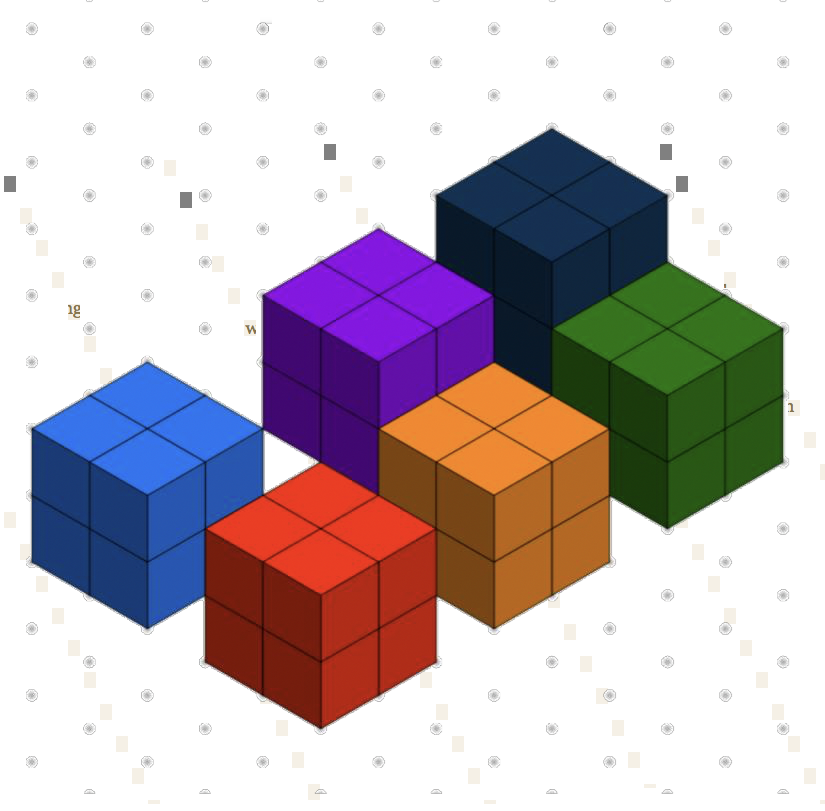}}
\hfill 
   \subfigure[Timing breakdown of the application of the Dirac
   operator for different lattice sizes.]{\includegraphics[width=.55\textwidth]{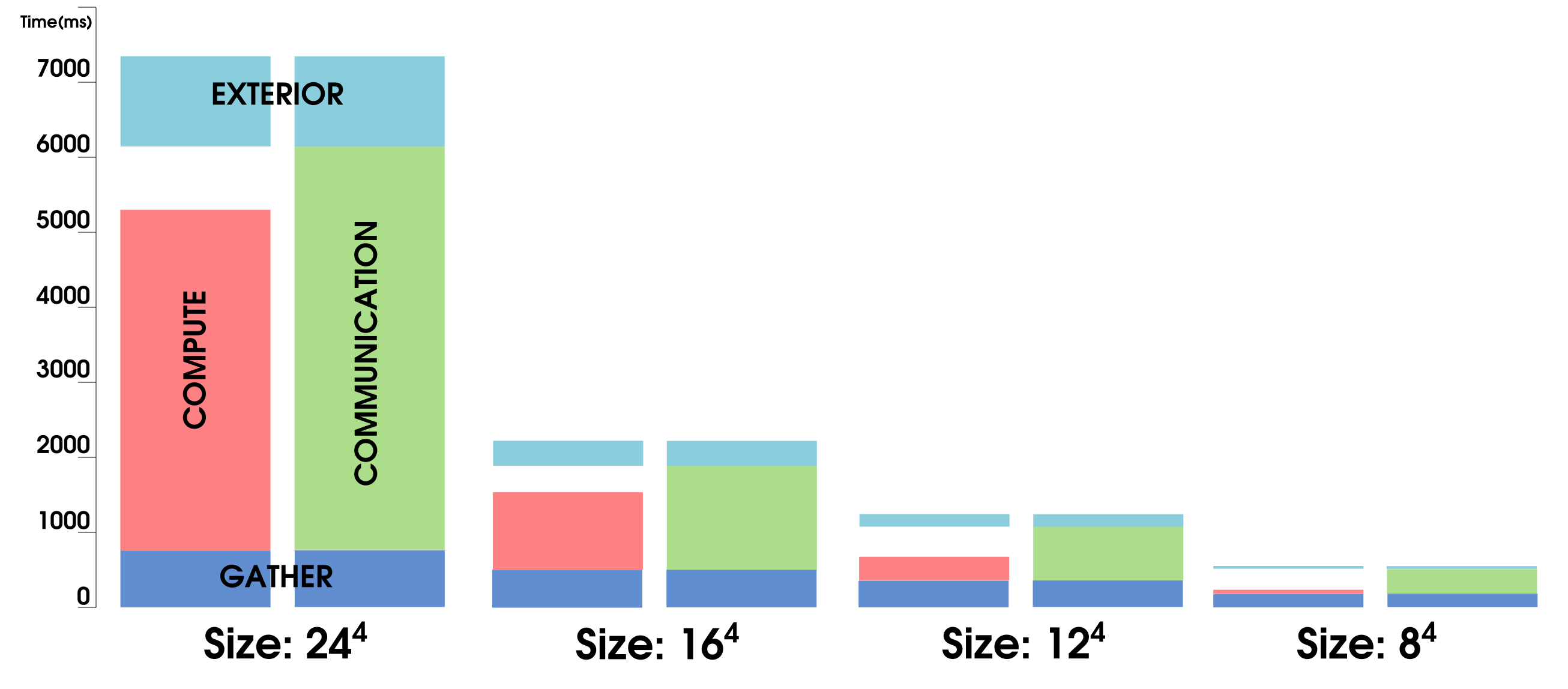}}
\hfill 
\caption{Grid MPI-3 shared memory communicators.}
  \label{network-grid-2}
\end{figure}

The authors have identified a possible workaround specific for Omni-Path:
they have used MPI-3 shared memory extension to accelerate intranode communications.
The MPI-3 shared memory features enable programmers to create regions of
shared memory that are accessible by the MPI processes. Hence the Grid
developers have used multiple MPI communications per node and
ensured that different ranks on the same node  assume
cartesian coordinates in cube of next-neighbours.
They differentiate between interior and external communication and use
for the former a direct copy into shared memory.
A breakdown of the computation and communications timings is reported in
figure \ref{network-grid-2}.b.
However communication overhead still dominates on small  local volumes. 

It must be noted that at present the Grid suite fully supports threaded MPI communications,
taking advantage of the latest driver release, an evolution that has drastically improved the
communication performances on the OPA hardware, compared to what was
reported during the conference\cite{grid-opa}.\\[.3cm]
{\bf Half Precision Communications} A more generic workaround takes
advantage of the tolerance that some inverter algorithms show against
the damaging of the neighbour information. C. Kelly and P. Boyle showed
that for a given inner conjugate gradient tolerance, matrix
inverter  precision can be kept roughly constant even if the neighbour
information is transmitted  only in half-precision.
The author have found a 45$\%$ reduction in communication time swapping to
 half precision communication. 
\begin{figure}[bht] 
  \centering
\includegraphics[width=.50\textwidth]{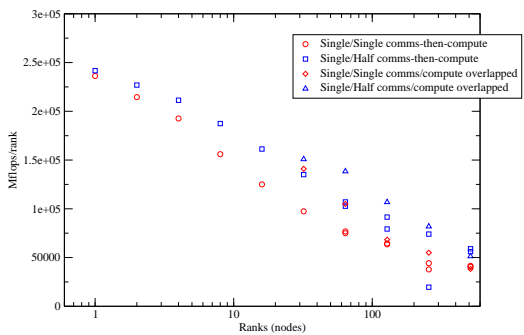}
  \caption{Strong-scaling performance of Grid halo exchange for single
    and half precision communication on Cori II.}
\label{network-half-prec-grid}
\end{figure} 
They also note that a careful
 implementation of the compress/decompress procedure was needed not to add any additional overhead.
The strong scaling investigation of this algorithm for the
halo-exchange of Grid is reported in figure \ref{network-half-prec-grid}, the labels single/single
and single/half refer to the overall floating point precision of the
Dslash and the precision of the communications.

\section{Conclusions}\label{end}
Data Motion, Memory and Cache bounds are defining the performances of
 QCD codes. Being capable to face them all at the same time is the
challenge that every lattice QCD programmer is facing nowadays. We are moving
towards a more hybrid kind of programming although there is always the
hope that CPU evolution will move along compiler's capability of
optimising code, e.g taking advantage in auto-vectorising large
vector units (AVX512).
GPU and in particular the QUDA libraries implementation have shown
very good performances, proving that the coding model of GPU is
portable over the hardware evolution with
little coding cost.
Concerning the new network hardware by Intel, all  concerns seem to be
software related and might have already been solved in the new version
of the PSM2 driver. Quite likely we will discover the new
performances in next year's report.\\[.3cm]
{\bf Acknowledgements}: I would like to thank the conference's
organizers for the invitation to present this review. I am indebted to Guido Cossu for
insightful comments at all stages of this work.  Acknowledgements for
support go to the STFC Consolidated Grants
ST/L000350/1 and  ST/P000479/1.

\clearpage
\bibliography{lattice2017}

\end{document}